\documentclass[fleqn,usenatbib]{mnras}

\usepackage{newtxtext,newtxmath,subfigure}

\usepackage[T1]{fontenc}

\DeclareRobustCommand{\VAN}[3]{#2}
\let\VANthebibliography\thebibliography
\def\thebibliography{\DeclareRobustCommand{\VAN}[3]{##3}\VANthebibliography}

\usepackage{graphicx}	
\usepackage{amsmath}	

\newcommand{\vasym}{\,$v_{\rm asym}$}
\makeatletter

\makeatletter

\makeatletter
\DeclareRobustCommand{\HI}{%
  \mbox{H\check@mathfonts\fontsize\sf@size\z@\selectfont I}%
}
\makeatother
\makeatletter
\DeclareRobustCommand{\HII}{%
  \mbox{H\check@mathfonts\fontsize\sf@size\z@\selectfont II}%
}
\makeatother

\title[Kinematic asymmetries in stars and gas]{The MAGPI Survey: Using kinematic asymmetries in stars and gas to dissect drivers of galaxy dynamical evolution}

\author[R. S. Bagge]{
R. S. Bagge,$^{1,2}$\thanks{E-mail: r.bagge@unsw.edu.au}
C. Foster,$^{1,2}$
F. D'Eugenio,$^{3,4}$
A. Battisti,$^{5,2}$
S. Bellstedt,$^{6}$
C. Derkenne,$^{9,2}$
S. Vaughan,$^{7,2}$
\newauthor
T. Mendel,$^{5,2}$
S. Barsanti,$^{5,2,7}$
K. E. Harborne,$^{6,2}$
S. M. Croom,$^{8,2}$
J. Bland-Hawthorn,$^{8,2}$
K. Grasha$^{5,2}$\thanks{ARC DECRA Fellow},
\newauthor
C. D. P. Lagos$^{6,2}$,
S. M. Sweet,$^{7,2}$
A. Mailvaganam,$^{9,2}$ 
T. Mukherjee,$^{9,2}$
L. M. Valenzuela,$^{10}$
J. van de Sande,$^{8,2}$
\newauthor
E. Wisnioski,$^{5,2}$
T. Zafar$^{9,2}$
\\
$^{1}$School of Physics, University of New South Wales, Kensington, NSW, 2032, Australia\\
$^{2}$ARC Centre of Excellence for All Sky Astrophysics in 3 Dimensions (ASTRO 3D)\\
$^{3}$Kavli Institute for Cosmology, University of Cambridge, Madingley Road, Cambridge, CB3 0HA, United Kingdom,\\
$^{4}$Cavendish Laboratory - Astrophysics Group, University of Cambridge, 19 JJ Thomson Avenue, Cambridge, CB3 0HE, United Kingdom,\\
$^{5}$Research School of Astronomy and Astrophysics, Australian National University, Cotter Road, Weston Creek, ACT, 2611, Australia\\
$^{6}$International Centre for Radio Astronomy Research, The University of Western Australia, 35 Stirling Highway, Crawley WA 6009, Australia,\\
$^{7}$School of Mathematics and Physics, University of Queensland, Brisbane, QLD 4072, Australia,\\
$^{8}$Sydney Institute for Astronomy, School of Physics, University of Sydney, NSW 2006, Australia,\\
$^{9}$School of Mathematical and Physical Sciences, Macquarie University, NSW 2109, Australia,\\
$^{10}$Universitäts-Sternwarte, Fakultät für Physik, Ludwig-Maximilians-Universität München, Scheinerstr. 1, 81679 München, Germany
}

\date{Accepted XXX. Received YYY; in original form ZZZ}

\pubyear{2023}

\begin{document}
\label{firstpage}
\pagerange{\pageref{firstpage}--\pageref{lastpage}}
\maketitle

\begin{abstract}
We present a study of kinematic asymmetries from the integral field spectroscopic surveys MAGPI and SAMI. By comparing the asymmetries in the ionsied gas and stars, we aim to disentangle the physical processes that contribute to kinematic disturbances. We normalise deviations from circular motion by $S_{05}$, allowing us to study kinematic asymmetries in the stars and gas, regardless of kinematic temperature. We find a similar distribution of stellar asymmetries in galaxies where we do and do not detect ionised gas, suggesting that whatever is driving the stellar asymmetries does not always lead to gas removal. In both MAGPI and SAMI, we find an anti-correlation between stellar asymmetry and stellar mass, that is absent in the gas asymmetries. After stellar mass and mean-stellar-age matching distributions, we find that at all stellar masses, MAGPI galaxies display larger stellar asymmetry compared to SAMI galaxies. In both MAGPI and SAMI galaxies, we find that star-forming galaxies with old mean-stellar-ages typically have larger asymmetries in their gas compared to their stars, whereas galaxies with young mean-stellar-ages have larger asymmetries in their stars compared to their gas. We suggest that this results from continuous, clumpy accretion of gas.
\end{abstract}




\section{Introduction}
In a Cold Dark Matter Universe with a cosmological constant ($\Lambda$CDM), dark matter haloes form from the collapse of primordial density perturbations \citep[e.g.,][]{2019Galax...7...81Z}. Baryons are dragged into these haloes by gravity, gaining angular momentum through tidal torques from surrounding halos \citep{1969ApJ...155..393P,1984ApJ...286...38W}. The baryons begin to cool, conserving angular momentum and falling further into the halo where they collapse and a rotating stellar disk begins to form \citep{1980MNRAS.193..189F,1998MNRAS.295..319M}.

This idea of galaxies beginning as rotating disk systems is supported by observations of galaxies having disks in place since $z>4$ \citep{2020Natur.581..269N,2021Sci...371..713L,2022ApJ...938L...2F,2023arXiv230814798T}, demonstrating significant rotation in their gas \citep{2021A&A...647A.194F,2023A&A...672A.106L} at $z\sim 3$, and stars at $z\sim 1$ \citep{2018ApJ...862..126N,2018ApJ...858...60B}. However, the prevalence of rotational support in galaxies seems to change over cosmic time; at low redshift, massive and quiescent (non-star-forming) galaxies are more dispersion dominated and less disk-like \citep{2007MNRAS.379..401E,2016ARA&A..54..597C} whereas at high redshift, star-forming galaxies exhibit higher gas turbulence in rotating disks compared to their low redshift counterparts \citep[characterised by smaller ratios of rotational to dispersion velocity, (V/$\sigma$);][]{2012ApJ...758..106K,2019ApJ...880...48U,2019ApJ...886..124W,2019MNRAS.490.3196P,2021MNRAS.507.3952R}. 

Massive galaxies in the local Universe are thought to have formed most of their stellar mass through dissipative gas cooling before and during Cosmic Noon \citep[$z\sim 2$;][]{2009Natur.457..451D,2014ARA&A..52..415M,2020ARA&A..58..661F,2023Sci...379.1323E}, becoming quiescent before continuing to grow in stellar mass from the accretion of lower mass, gas-poor satellites \citep{2009ApJ...699L.178N,2010ApJ...725.2312O,2014ApJ...788...28V,2021ApJ...908..135S}. Accretion is not only restricted to stars; galaxies can acquire new gas from gas-rich mergers \citep[e.g.,][]{2014A&A...567A..68D}, along cold gas filaments \cite[e.g.,][]{2009MNRAS.395..160K}, or from the hot outer halo \citep[e.g.,][]{2015MNRAS.448.1271L}. Galaxies can also re-accrete gas that has been expelled from their disk through `galactic fountain' effects \citep{1976ApJ...205..762S, 2008A&A...484..743S,2010MNRAS.404.1464M}. Indeed, some form of new gas accretion has to be invoked to explain the Star Formation Rates (SFRs) and cold gas masses seen at high redshift \citep{2016ApJ...826..214B,2017ApJ...837..150S,2018ApJ...853..179T} as well as the growth of stellar mass over cosmic time \citep[e.g.,][]{2020ApJ...902..111W}

The angular momentum of this accreted material is not necessarily aligned with the spin axis of the existing stellar disk, consequently, if this newly accreted gas can form stars, there may be a kinematic misalignment, (i.e., an offset between the kinematic position angles of the ionised gas and stars). This transitory state is sometimes observed at low redshift \citep{2019MNRAS.483..458B}. If the accretion rate of this event is sufficiently low, it is unlikely to cause a significant disturbance in the gas \textit{or} stellar kinematics, since the existing stellar disk will provide a torque on the accreted gas, aligning it into a similar configuration as the stars \citep[this includes counter-rotating disks, e.g.,][]{2006MNRAS.373..906M, 2017MNRAS.471L..87O,2019MNRAS.483..458B,2019MNRAS.483..172D}. But if the accreted gas came in a `clumpy' fashion (i.e. a merger event), or if there was very little gas present in the galaxy, to begin with, there would be significant disturbances in ionised gas kinematics. This phenomenon has been found in both high and low redshift merging galaxies \citep{2008ApJ...682..231S,2018MNRAS.476.2339B,2020ApJ...892L..20F}. Along with heavily disturbing the ionised gas, the stellar kinematics also become disturbed in merging galaxies \citep{2011MNRAS.416.1654B,2022MNRAS.509.4372L,2023arXiv230305520D}.

Signatures of disturbed kinematics in the stars and ionised gas can be quantified in different ways. Commonly for spatially resolved spectroscopy surveys (radio and optical), this is done by computing deviations from circular motion that are captured by higher-order terms in a Fourier Series fitted to concentric ellipses (or projected tilted rings) on a velocity map \citep{2008MNRAS.390...93K,2011MNRAS.414.2923K,2021ApJ...923..220D,2023arXiv230502959G}. Disturbances to kinematics will cause the velocity profile around these ellipses to appear asymmetric, and these `asymmetries' can be quantified by how much power is in the fitted higher order Fourier coefficients \citep[e.g.,][]{2006MNRAS.366..787K}. Velocity asymmetries in the ionised gas and stars have been linked to aspects of galaxy evolution such as non-axisymmetric disk features (e.g., bars), stellar mass and local environment \citep{2014A&A...568A..70B, 2017MNRAS.465..123B,2018MNRAS.476.2339B,2021MNRAS.505.3078V}.

Gas and stars can be thought of as fundamentally different fluids; with gas considered collisional, whereas stars are collisionless \citep{1984ApJ...276..114J}. By being a collisional fluid, gas within a galaxy will be able to correct from induced kinematic disturbances faster than stars would be able to. As such, if we were to compare disturbances in both the stars and gas, we would expect the disturbance to last in the stellar kinematics, possibly long after the gas has resettled \citep[e.g.,][]{1962ApJ...136..748E}. Comparison of gas and stellar disturbances can be used to isolate specific drivers behind kinematic asymmetries, which can be difficult to discern when only considering the ionised gas \citep{2023arXiv231110268B}, or even different gas phases \citep{2023MNRAS.519.1452W}.

In this work, we study the source of kinematic disturbances in galaxies by comparing the kinematic asymmetries between stars and ionised gas to disentangle what is driving kinematic asymmetries in each. We also look for evidence of dynamical evolution in these asymmetries at different redshifts. To this end, we draw a sample of galaxies from the Middle Ages Galaxy Properties in IFS \citep[MAGPI;][]{2021PASA...38...31F} and the SAMI Galaxy Survey Data Release 3 \citep{2021MNRAS.505..991C}. We use \textsc{kinemetry} \citep{2006MNRAS.366..787K} to model the stellar and ionised gas velocity maps and use the asymmetry to measure the amount of disturbance in each. The structure of this paper is as follows; \S\ref{MAGPI} gives a brief overview of the MAGPI survey and data products we use in this work, \S\ref{sample_select} describes our method of sample selection in both the MAGPI and SAMI surveys, \S\ref{kinemetry} describes how we use \textsc{kinemetry} in our analysis of the stellar and ionised gas velocity maps. In \S\ref{Res}, we show our results and discuss possible interpretations, before offering our conclusions in \S\ref{Conc}. We adopt a \cite{2020A&A...641A...6P} cosmology for this work; explicitly a flat Universe with $H_0$ = 67.7 kms$^{-1}$ Mpc$^{-1}$, $\Omega_{\rm M}$ = 0.31 and $\Omega_\Lambda$ = 0.69. All correlations quoted are measured using Pearson correlation coefficients.

\section{Data}
\subsection{The MAGPI Survey}
\begin{figure*}
    \centering
    \includegraphics[scale=0.55]{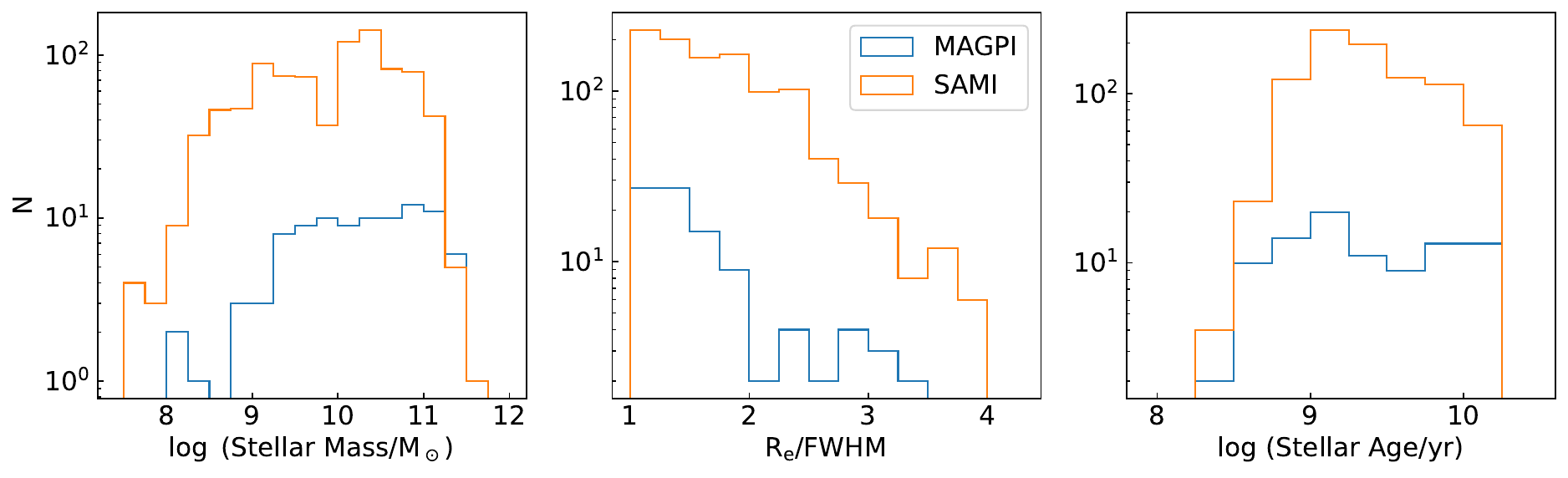}
    \caption{Histograms of the mass (left), R$_{\rm e}$/FWHM (middle) and mean stellar population age (right) distributions for our MAGPI (blue) and SAMI (orange) subsamples. The bin sizes for stellar mass and mean stellar population age are 0.25 dex, and the bins sizes for R$_{\rm e}$/FWHM is 0.25. Our MAGPI subsample is skewed towards more massive galaxies, while our R$_{\rm e}$/FWHM and mean stellar population age distributions are similar.}
    \label{MAGPI_SAMI_hist}
\end{figure*}
The MAGPI survey is an ongoing MUSE/VLT Large Program (Program ID: 1104.B-0536) targeting 56 (42 completed as of January 2024) 1 $\times$ 1 arcmin fields from the Galaxy and Mass Assembly \citep[GAMA;][]{2011MNRAS.413..971D} fields G12, G15 and G23. MAGPI is currently collecting spatially resolved spectra of stars and ionised gas at a redshift of 0.28 $\leq z <$ 0.35. MAGPI also includes archival observations of legacy fields Abell 370 and Abell 2744. When observations are completed, MAGPI's entire dataset will include 60 spatially resolved `primary' targets (M$_*>$ 7 $\times$ 10$^{10}$ M$_\odot$) and $\sim$ 100 spatially resolved `secondary' targets (M$_*\geq$ 10$^{9}$ M$_\odot$) at $z\sim$ 0.3.

Data was collected using MUSE Wide Field Mode in the nominal wavelength range (4650 \AA\ - 9300 \AA) with a spectral sampling of 1.25 \AA. Ground Level Adaptive Optics (GLAO) are employed to mitigate the effect of atmospheric seeing, resulting in a 270 \AA-wide gap between 5780 \AA\ and 6050 \AA\ due to the GALACSI laser notch filter. Each MAGPI field has a field-of-view (FOV) equal to 1$\times$1 square arcmin with spatial sampling of 0.2$''$ pixel$^{-1}$ and average image quality of 0.65$''$. A detailed discussion of the data reduction will be presented in the first data release (Mendel et. al., in prep).

For this work, stellar masses are derived using the SED-fitting code \textsc{ProSpect} \citep{2020MNRAS.495..905R} with 9-band $u$-$K_s$ (KiDS and VIKING imaging) photometry pixel-matched to the MAGPI cubes, where forced photometry is based on the MAGPI segmentation maps. The implementation of \textsc{ProSpect} applied is consistent with the method used to derive stellar masses for GAMA \citep{2020MNRAS.498.5581B,2022MNRAS.513..439D}. A truncated skewed-normal parametrisation is used for the star-formation history, with a linearly evolving metallicity evolution implemented to ensure a chemical build-up that follows the stellar mass build-up in the galaxy. A \cite{2003PASP..115..763C} Initial Mass Function (IMF) is assumed, and the \cite{2003MNRAS.344.1000B} stellar population templates are used. 

Stellar kinematic maps for MAGPI galaxies are obtained using \textsc{pPXF}, which models the stellar continuum as a linear superposition of input stellar templates. A subset of the reduced IndoUS stellar-template library \citep{2004ApJS..152..251V} is used. Detailed description is provided in \citep{2023arXiv230304157D}. Briefly, all stellar kinematics are obtained by fitting their stellar templates on a spaxel-by-spaxel basis using additive Legendre polynomials of degree $n$ = 12. Regions of possible nebular emission and the strongest sky-lines are masked. Additionally, and for MAGPI only, the spectral region affected by the AO laser is also masked. A first round of fitting is completed on a set of elliptical annuli to derive a subset of optimal stellar templates to be used to fit the spectrum of individual spaxels. Starting from the centre, the procedure creates a series of elliptical annuli, using only spaxels with nominal SNR $\geq$ 3. Each annulus is grown in steps of one half-pixel (0.1$''$) until either the target SNR = 25 is (nominally) reached, or until there are no more valid spaxels to merge. A minimum of one (elliptical) bin is required to proceed; galaxies with insufficient SNR are immediately failed. The procedure then fits the stellar kinematics of each spaxel, by restricting the pool of input templates to the subset used to create the best-fit model spectrum of the annular bin(s) that the current spaxel intersects, plus one more external annulus bin (if present) and one more internal annulus (if present).

$\lambda_{\rm Re}$ is the spin parameter proxy defined by \cite{2007MNRAS.379..401E}. Explicitly, $\lambda_{\rm Re}$ is, 

\begin{equation}
    \lambda_{R\rm e} = \frac{\sum_{i=1}^N F_i R_i |V_i|}{\sum_{i=1}^N F_i R_i \sqrt{V_i^2 + \sigma_i^2}}
    \label{lambda_RE}
\end{equation}

where $N$ is number of spaxels in the chosen aperture, $F_i$ is the flux within the $i$-th spaxel, $V_i$ is the line-of-sight velocity, $\sigma_i$ is the line-of-sight velocity dispersion, and $R_i$ is the galactocentric radius to the $i$-th spaxel. The computation of $\lambda_{R\rm e}$ will be described in detail in Derkenne et al., in prep., but are briefly described here. Apertures are first constructed using a Multi-Gaussian Expansion \citep{2002MNRAS.333..400C} fit to a mock $r$-band image of the galaxy. Using that aperture, $\lambda_{R \rm e}$ is computed following Eqn. \ref{lambda_RE} and corrected for seeing following \cite{2020MNRAS.497.2018H}.

Stellar ages are computed using \textsc{pPXF} to fit MILES single stellar population templates to 1$R_e$ integrated spectra with emission lines included in the fit. We use light-weighted stellar ages, as opposed to mass-weighted stellar ages.

Ionised gas velocity maps are obtained from continuum-subtracted spectra using GIST (Galaxy IFU Spectroscopy Tool; \citealt{2019A&A...628A.117B}), a \textsc{python} wrapper of \textsc{pPXF} \citep{2004PASP..116..138C,2017MNRAS.466..798C} and \textsc{GandALF} \citep{2006MNRAS.366.1151S,2006MNRAS.369..529F}. Briefly, a set of 22 single-component Gaussians are fit to spectral lines ranging from [O~{\sc ii}]\,3727\AA\ to [S~{\sc ii}]\,6733\,\AA\ where the widths and relative velocities are all tied to one line (see \cite{2021PASA...38...31F} for more details). A detailed description of the emission line products will be present in Battisti et al., in prep. We use [S~{\sc ii}]\,6718,33\,\AA\ integrated fluxes within 1$R_e$ to classify each of the galaxies in our sample into star-forming and AGN galaxies. 

\subsection{The SAMI Galaxy Survey}
\label{MAGPI}
The Sydney-AAO Multi Integral field (SAMI) Galaxy Survey is an integral field spectroscopic survey of galaxies within the local Universe ($0.004< z <0.095$), probing galaxies over a large stellar mass range of $10^7<M_*<10^{12}$ M$_\odot$, morphology and galactic environments \citep{2015MNRAS.447.2857B, 2017MNRAS.468.1824O,2017ApJ...844...59B}. Spectral cubes were obtained using the SAMI instrument \citep{2012MNRAS.421..872C} on the 3.9m Anglo-Australian Telescope connected to the AAOmega spectrograph \citep{2006SPIE.6269E..0GS}. The SAMI instrument has 13 hexabundles and 26 sky fibres over a 1-degree field-of-view. Each hexabundle consists of 61 fibers with a diameter of 1.6 arcsec, giving each hexabundle a diameter of 15 arcsec. Data products used in this work (gas velocity maps, emission line fluxes, stellar populations), barring the stellar masses, are computed similarly to the MAGPI data products previously discussed and are described, in detail, in the SAMI Galaxy Survey Data Release papers \citep{2018MNRAS.475..716G,2018MNRAS.481.2299S,2021MNRAS.505..991C}. Stellar masses for SAMI galaxies are computed using extinction-corrected $g-i$ colors from the GAMA catalogue, following Eqn. 3 in \cite{2015MNRAS.447.2857B}. SAMI stellar kinematics are derived in a similar manner to aforementioned MAGPI stellar kinematics and a detailed description is provided in \cite{2017ApJ...835..104V}.

\begin{figure*}
    \centering
    \includegraphics[scale=0.55]{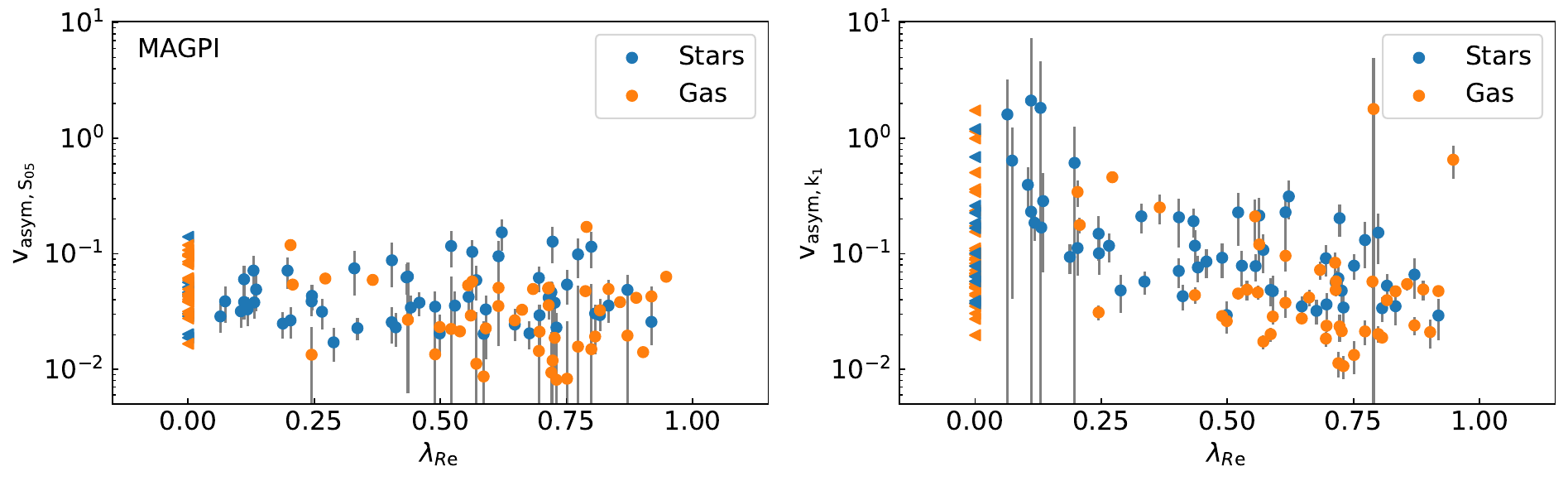}
    \caption{\vasym\ against $\lambda_R$ for MAGPI galaxies. \vasym\ normalised by $S_{05}$ is shown on the left, while \vasym\ normalised by $k_1$ is shown on the right. \vasym\ measured in the stars is shown in blue, while \vasym\ measured in the gas is in orange. Stellar non-detections are shown as orange triangles. When normalising by $S_{05}$, we see that \vasym\ is largely independent of whether the galaxy is rotating slowly (small $\lambda_{R\rm e}$) or fast (large $\lambda_{\rm Re}$), whereas if we normalise by $k_1$, the \vasym\ value, for slow rotating galaxies begins to diverge. 7 galaxies are not shown due to no $\lambda_{R\rm e}$ values being available, and 16 galaxies are not shown due to the measured $\lambda_{R\rm e}$ being unphysical (i.e., $\lambda_{R\rm e}>1$). These galaxies would be considered fast rotating galaxies ($\lambda_{R\rm e}$>0.6) prior to seeing correction.}
    \label{k1_vs_s05}
\end{figure*}

\subsection{Sample Selection}
\label{sample_select}
To select MAGPI galaxies, we only select those that have $R_{\rm e}$/FWHM $>$ 1 and are at a redshift of $0.2<z<0.4$. We further select those galaxies where the minimum H$\alpha$ SNR/spaxel $>$20 and stellar continuum SNR/spaxel $>$3 within 2$R_{e}$. Galaxies that met the stellar continuum criterion, but not the H$\alpha$ criterion are included as ionised gas non-detections. Similarly, galaxies that met the H$\alpha$ criterion, but not the stellar continuum criterion are included as stellar non-detections. Finally, we use [S~{\sc ii}]-BPT diagnostic plot with integrated $1R_{\rm e}$ fluxes to classify our sample into star-forming and AGN galaxies following the criteria stipulated in \citep{2006MNRAS.372..961K}. The [S~{\sc ii}]-BPT was used as it allows a cleaner separation between star-forming, AGN and LINER galaxies, than the [N~{\sc ii}]-BPT alone \citep{2001ApJ...556..121K,2016MNRAS.461.3111B,2021ApJ...915...35L}. Galaxies that host AGN/LINER are removed from the sample. This yields a MAGPI subsample of 95 galaxies. 

To select SAMI galaxies, we employ the same SNR, spatial resolution ($R_{\rm e}$/FWHM$>$1), and ionisation source requirements on galaxies in SAMI Data Release 3 \citep{2021MNRAS.505..991C}. We do not place any redshift selection for SAMI galaxies, for which the median redshift is 0.04. As of writing, MAGPI does not include any cluster galaxies in its master catalogue, because of this we exclude cluster galaxies within SAMI from the analysis. Our final subsample contains 802 galaxies. 

Distributions of the stellar mass, $R_{\rm e}/$FWHM and mean-stellar-ages are shown in Fig. \ref{MAGPI_SAMI_hist}. Due to the large difference in sample sizes between SAMI and MAGPI, we seperately match the stellar mass, $R_{\rm e}$/FWHM and mean stellar population age distributions between surveys, and discuss this in Sect. \ref{dynamical_evolution}. The median seeing for our SAMI subsample is coarser compared to our MAGPI subsample, $\sim 2''$ versus $\sim 0.7''$, respectively. These seeing conditions correspond to a similar physical spatial resolution of $\sim 3$ kpc for MAGPI, and $\sim 1.5$ kpc for SAMI.

\subsection{Kinemetric Analysis}
\textsc{kinemetry} \citep{2006MNRAS.366..787K} is an algorithm that extends photometric analysis commonly used in surface brightness photometry to higher order moments of the line-of-sight-velocity-distribution \citep[e.g., mean velocity, dispersion;][]{1993MNRAS.265..213G,1993ApJ...407..525V}. \textsc{kinemetry} constructs models of the line-of-sight-velocity-distribution (LOSVD) moments by fitting a series of concentric ellipses with position angle (PA) and axial ratio $q=b/a$, where $b$ and $a$ are the semi-minor and semi-major axes, respectively. Like other tilted-ring fitting algorithms \citep{1974ApJ...193..309R,1987hrcs.book.....B,2007ApJ...664..204S,2015MNRAS.451.3021D}, \textsc{kinemetry} fits a Fourier Series along the ellipse, 
\begin{equation}
    K(a,\theta) = A_0 + \sum_{m=1}^{m=N} A_m\sin{(m\theta)} + B_m\cos{(m\theta)},
    \label{FS}
\end{equation}
where $a$ is the semi-major axis of the ellipse, $\theta$ is the azimuth along the ellipse with respect to the semi-major axis, A$_0$ is the zeroth harmonic term and A$_m$ and B$_m$ are the $m\rm{th}$ additional harmonic terms. \textsc{kinemetry} determines the best fitting PA and $q$ by minimising the harmonic coefficients that are in the series. Eqn. \ref{FS} can be represented more compactly as,  
\begin{equation}
    K(a,\theta) = A_0 + \sum_{m=1}^{m=N} k_m\cos{(m[\theta - \phi_m(a)])},
    \label{compactFS}
\end{equation}
where $k_m = \sqrt{A_m^2+B_m^2}$ and $\phi_n = \arctan\frac{A_m}{B_m}$. The $k_m$ parameters describe the kinematics of the galaxy while PA and $q$ describe the geometry of the ellipse. A thin disk galaxy rotating with perfect circular motion can be described with a single cosine term in Eqn. ~\ref{FS} (i.e., $V(\theta)=B_1\cos{\theta}$) which is a symmetrical function of azimuth around $\theta$=180. Any deviation from non-circular motion results in an asymmetric function, and these asymmetries are encoded as power to higher-order coefficients. If one used \textsc{kinemetry} on a velocity map (1$^{\rm st}$ moment), $k_1$ would represent the rotational velocity, $V_{\rm rot}$ at that ellipse. In the instance one used \textsc{kinemetry} on a dispersion map (2$^{\rm nd}$ moment), $A_0$, would represent the $\sigma$ at that ellipse.

\subsubsection{Defining kinematic asymmetry}
In the literature, asymmetry in velocity maps for galaxies is usually defined as the higher order (up to $n=5$) coefficients in Eqn. ~\ref{compactFS}, normalised to $k_1$ \citep{2011MNRAS.414.2923K,2017MNRAS.465..123B,2020ApJ...892L..20F}. For this study, we use a different formalism, where we normalise the higher order coefficients with the $S_{05}$ parameter instead where $S_{05} = \sqrt{0.5V_{\rm rot}^2 + \sigma^2}$, \citep{2006ApJ...653.1027W,2006ApJ...653.1049W,2014ApJ...795L..37C,2019MNRAS.487.2924B}. Where $V_{\rm rot}$ is the rotational velocity taken where the rotational curve reaches its maximum, and $\sigma$ is the light-weighted average dispersion within 1$R_{\rm e}$. $V_{\rm rot}$ is corrected for inclination by V$_{\rm obs}$/$\sin{i}$, where $V_{\rm obs}$ is the fitted $k_1$ value, and $i$ is inclination of the galaxy.

Explicitly, the new asymmetry parameter is,
\begin{equation}
    v_{\rm asym} = \frac{k_2+k_3+k_4+k_5}{4S_{05}}
\end{equation}
$S_{05}$ is calculated for both the stars and ionised gas and used separately. Both $V_{\rm rot}$ and $\sigma$ are measured using \textsc{kinemetry} where we fix the PA and $q$ for all radii to their $i$-band photometric values. 

We now discuss our motivation for changing our asymmetry parameter. If we normalise the higher order coefficients by $k_1$ measured in galaxies that are slowly rotating, not rotating at all, or are dispersion dominated we would be dividing by a small number (i.e., $k_1$ $\rightarrow$ 0). This would artificially enhance \vasym, which would not necessarily be a true indication of how disturbed the kinematics are in those galaxies. To avoid this issue, $k_1$ (i.e., $V_{\rm rot}$) is removed and replaced by $S_{05}$, which has been shown to have a tight relation with stellar mass in galaxies of all Hubble types \citep[e.g., Es, Sp, S0s;][]{2014ApJ...795L..37C} and be an effective tracer of gravitational potential well depth \citep{2012ApJ...758..106K}. Including $\sigma$ in the normalisation of the higher order terms captures the full kinematic `budget' of the galaxy, and prevents the asymmetry value from diverging in slower rotating or dispersion-dominated galaxies. We discuss the impact of changing this normalisation in Sect. \ref{k1_s05}.

When measuring the kinematic asymmetries for both MAGPI and SAMI galaxies, we run \textsc{kinemetry} on the stellar and ionised gas velocity maps where kinematic position angle is allowed to vary between 0$^\circ$ and 360$^\circ$. Kinematic axial ratio $q_{\rm kin}$ is set to vary between $q_{\rm phot}\pm$0.1. We opted for tight limits on $q$ given our spatial resolution and not selecting galaxies on inclination
(See Sect. 2.2 of \citealt{2008ApJ...682..231S} and Sect. 4.2 of \citealt{2017ApJ...835..104V}). Spaxels on the stellar velocity maps with SNR/spaxel$<$3 are masked. A similar criterion is used for ionsied gas velocity maps where spaxels with SNR (H$\alpha$)/spaxel<3 are masked. We run \textsc{kinemetry} out to $1R_e$ and take the measured $v_{\rm asym}$ at 1$R_{\rm e}$. For MAGPI and SAMI galaxies, the distance between each ellipse is set to half the estimated seeing, $\sim0.3''$ and $\sim 1''$, respectively. Uncertainties for \vasym\ are estimated using 100 Monte Carlo realisations, where we re-run \textsc{kinemetry} on the velocity map with Gaussian noise injected into each spaxel corresponding to the error in the velocity measurements. The final \vasym\ values and uncertainties correspond to the mean and standard deviation of the Monte Carlo distribution, respectively.

\subsubsection{Impact of seeing on $\lambda_{R \rm e}$ and \textsc{kinemetry} asymmetries}
\label{seeing}
Galaxy kinematic measurements can be negatively impacted by atmospheric seeing conditions, where  poor seeing conditions can decrease the measured rotation amplitude, while increasing the measured velocity dispersion. This can prove to be problematic when investigating slow and fast rotator fractions in IFS surveys, and corrections are necessary to ensure $\lambda_{R\rm e}$ is correctly measured in fast-rotating galaxies \citep{2017ApJ...835..104V,2018ApJ...852...36G,2018MNRAS.477.4711G}. The $\lambda_{R\rm e}$ values used in this work are corrected using the empirical relation in \cite{2020MNRAS.497.2018H}, which considers how resolved the galaxy is ($\sigma_{\rm PSF}$/R$_{\rm e})$, its S\'{e}rsic index $n$, and ellipticity $\epsilon$ (Derkenne et al., in prep.). The mean amplitude of this corrections is 0.18 and standard deviation of 0.16.

In contrast, the effect of seeing on kinematic asymmetries is such that as the seeing degrades, complex kinematic features progressively are washed out and velocity fields become more regular, decreasing the asymmetry. This effect was investigated in App 2.2 of \cite{2017ApJ...835..104V}, who degraded ATLAS$^{\rm 3D}$ kinematics maps to see how recoverable the kinematic asymmetry was. The authors report that poor seeing reduced the kinematic asymmetry in those galaxies that had large kinematic asymmetry, but found little correlation between this impact and angular size of the galaxies. So while the complex kinematic features in galaxies that are very resolved are likely retained, these features are harder to measure in less resolved galaxies, leading to an underestimate of the true asymmetry.

However, a correction similar to that applied for $\lambda_{R\rm e}$ in fast-rotating galaxies cannot be performed since we do not know \textit{a-priori} if a low-asymmetry galaxy is intrinsically a high-asymmetry galaxy that has had its asymmetry washed out by seeing, or if it has intrinsically low-asymmetry. We plot $v_{\rm asym}$ against $R_{\rm e}/$FWHM for MAGPI and SAMI galaxies in Fig. \ref{vasym_re_over_psf}, and find no trend. The highest correlation coefficient is $\rho$=-0.23 with a p-value of 0.09 for stellar asymmetry. We also measure how \vasym\ varies as a function of the number of pixels that are along the ellipse. Again, we find only a weak correlation with the highest correlation coefficient being $\rho$=-0.20 with a p-value of 0.03. We leave this as caveat of our analysis.


\begin{figure}
    \centering
    \includegraphics[scale=0.5]{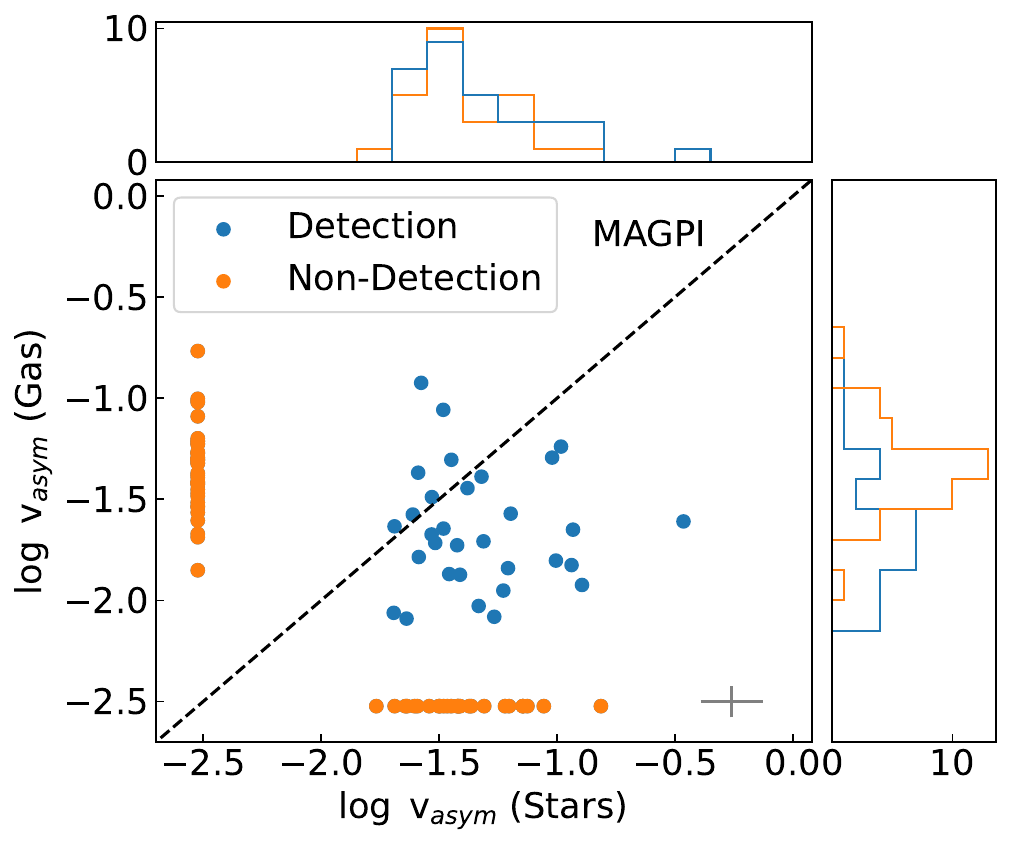}
    \caption{\vasym\ measured in the stars against \vasym\  measured in the ionised gas for MAGPI galaxies. The histograms above and to the right show the distributions of detections and non-detections in the ionised gas and stars, respectively. Galaxies with non-detections in either stars or gas are shown in orange. The black dot-dashed line is the 1:1 to delineate between galaxies with larger gas or stellar kinematic asymmetries. The average size of the error bars is shown in the bottom right corner. The distributions of stellar asymmetries with gas detections and gas non-detections are similar. If interactions and mergers are driving the stellar asymmetries, the similar distribution between the two suggests that interactions or mergers do not always remove the gas in galaxies.}
    \label{v_asym_sg_marginal}
\end{figure}

\begin{figure*}
    \centering
    \includegraphics[scale=0.6]{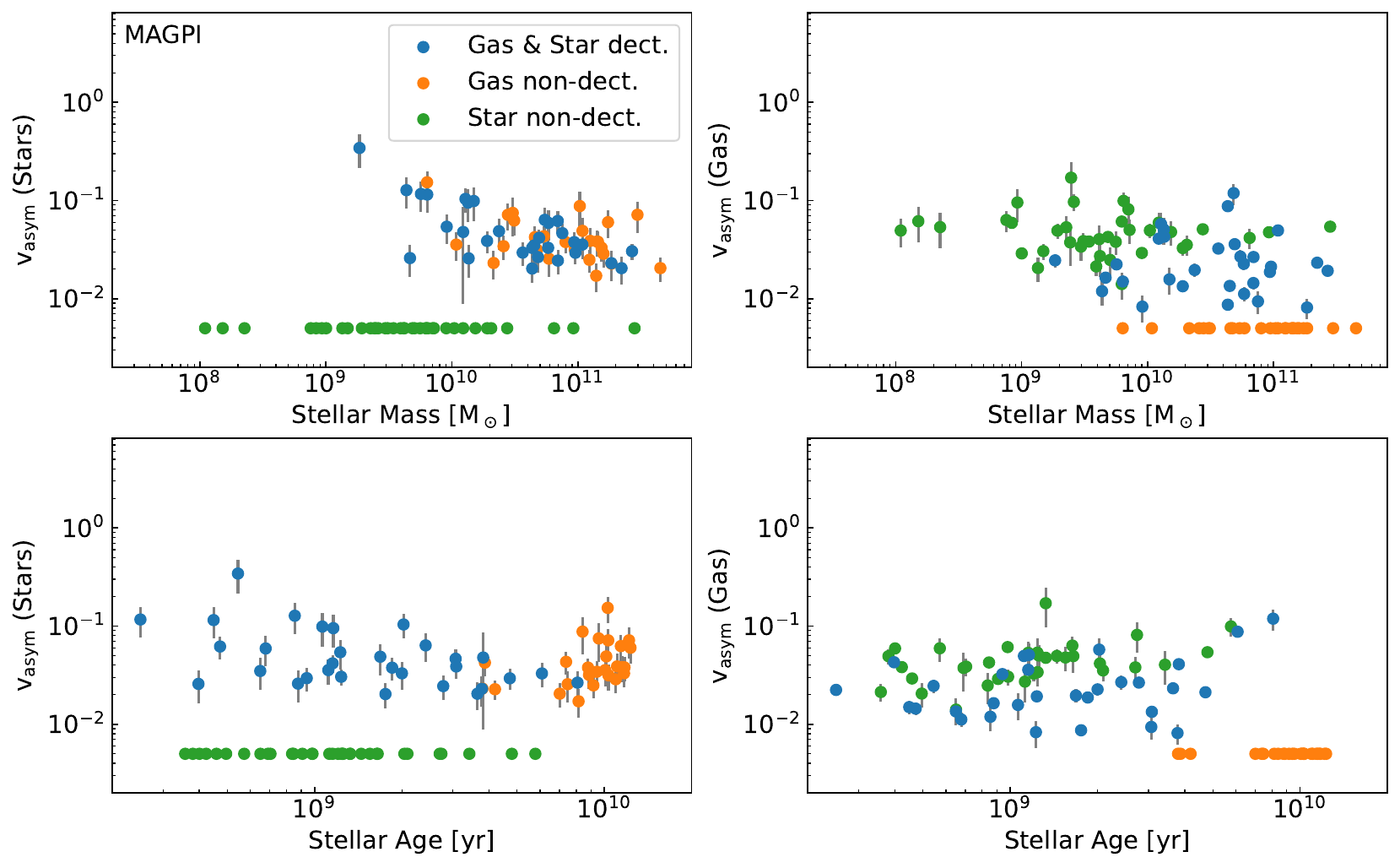}
    \caption{For MAGPI galaxies, stellar \vasym\ against stellar mass (\textit{top left}), ionised gas \vasym\ against stellar mass (\textit{top right}), stellar \vasym\ against stellar age (\textit{bottom left}), and ionised gas \vasym\ against stellar age (\textit{bottom right}). Ionised gas and stellar non-detections are shown in green and orange, respectively. There is a anti-correlation between stellar asymmetries and stellar mass ($\rho$=-0.51,p<10$^{-9}$), that is much weaker ($\rho$=0.29,p=0.29), and more scattered in the ionised gas. We interpret this as the gas `washing out' asymmetries faster than the stars, a consequence of the collisional nature inherent to the gas.}
    \label{asym_stellar_mass}
\end{figure*}

\section{Results and Discussion}
\label{Res}
In this section we compare \vasym\ in the stars and the ionised gas that was normalised by $k_1$ and normalised by $S_{05}$, discuss the physical drivers behind the asymmetries in the stars and ionised gas for MAGPI galaxies and the possible dynamical evolution of galaxies from MAGPI to SAMI.

\subsection{Stellar and Ionised Gas Kinematic Asymmetries for MAGPI Galaxies}
\subsubsection{Normalisation}
\label{kinemetry}
\label{k1_s05}
Fig. \ref{k1_vs_s05} shows the \vasym\ measured in the stars and ionised gas against $\lambda_R$. We see that when normalising by $k_1$, galaxies will typically exhibit larger asymmetry values with comparatively enhanced scatter at low $\lambda_{R\rm e}$ (i.e., slower rotating galaxies). Conversely, when normalising by $S_{05}$, \vasym\ appears largely independent of $\lambda_{R\rm e}$. This confirms our hypothesis that changing the normalisation will minimise diverging for slower rotating systems. Adopting \vasym\ to be $v_{\rm{asym}, S_{05}}$ we show in Fig. \ref{v_asym_sg_marginal} \vasym\ measured in stars against \vasym\ measured in the gas. There is no clear correlation between asymmetry in the stars and asymmetry in the gas, suggesting that whatever is driving the asymmetry in the stars and gas is not coupled. The lack of a correlation could also be due to the gas being able to return to low asymmetry through collisional processes, which stars cannot do. We find that most of our sample demonstrates larger stellar asymmetries (i.e., are below the dashed line). This is not unexpected, as stellar velocity disturbances could remain present in the kinematics for a longer time, compared to velocity disturbances in the gas. The distribution of stellar asymmetries between galaxies detected in ionised gas and those where we do not detect ionised gas is also similar, albeit with tail towards large stellar asymmetries. A Kolmogorov–Smirnov (KS)-test suggests that they could be different distributions ($t$=0.49, p=0.0001).

\subsection{Physical Drivers of Kinematic Asymmetries in MAGPI Galaxies}
\subsubsection{Stellar Mass and Gravitational Potential}
\label{mass}


Higher stellar mass galaxies will have a larger gravitational potential, and be able to provide a larger restoring force to kinematic disturbances. Since both stars and gas respond to the same gravitational potential of the galaxy, both stellar and gas asymmetries should exhibit an anti-correlation with stellar mass since $S_{05}$ traces this potential, regardless if it was measured in the stars or gas. Fig. \ref{asym_stellar_mass} shows the asymmetries in the stars and gas plotted against stellar mass. Including non-detections, there is a strong anti-correlation between stellar mass and stellar asymmetry ($\rho$=-0.59,p<10$^{-5}$); however this anti-correlation is weaker when considering the gas asymmetries in both the stellar detections and non-detections ($\rho$=-0.30,p=0.006), and is entirely absent when considering the stellar detections and non-detections separately. Excluding gas non-detections increases the strength of the correlation between stellar asymmetries and stellar mass ($\rho$=-0.69,p<10$^{-5}$). 


The absence of a correlation between gas asymmetries and stellar mass is contrary to previous findings in the literature \citep{2017MNRAS.465..123B,2022ApJS..262....6F,2023arXiv231110268B}. These asymmetries, however, were normalised to $k_1$, not $S_{05}$, hence it is likely these lower mass galaxies were slower rotating, causing the \vasym\ to diverge. Normalising by $S_{05}$ should not remove a physical anti-correlation between dynamical disturbances and mass, as a tight $S_{05}$ vs. stellar mass correlation has been shown to be independent of whether $V_{\rm rot}$ and $\sigma$ are measured in the stars or gas \citep[][but see \citealt{2019MNRAS.487.2924B,2020MNRAS.498.5885B}]{2014ApJ...795L..37C}. For completeness, we check this by calculating $S_{05,\rm stars}$/$S_{05,\rm gas}$ for each galaxy, and find that this value is close to 1 across the sample, with a mean of 0.87 and standard deviation of 0.27. We conclude that this absence of an anti-correlation between \vasym\ in the gas and stellar mass is not an artefact of normalising by $S_{05}$ measured in different phases.

The lack of a correlation between stellar mass and ionised gas asymmetries suggests that gas can recover from induced asymmetries independent of the restoring force provided by the gravitational potential. This is possibly a result of ionised gas being able to dissipate energy and transfer angular momentum faster than the stars are able to through collisional processes. The additional restoring pathways available to the gas mean the asymmetries in the gas may be dissipated on shorter timescales, while still persisting in the stars. However, this is assuming that the gas and stellar asymmetries were induced in the galaxy at the same time, which we can not confirm given there is little correlation between the \vasym\ in the stars and \vasym\ in the gas.

\begin{figure}
    \centering
    \includegraphics[scale=0.55]{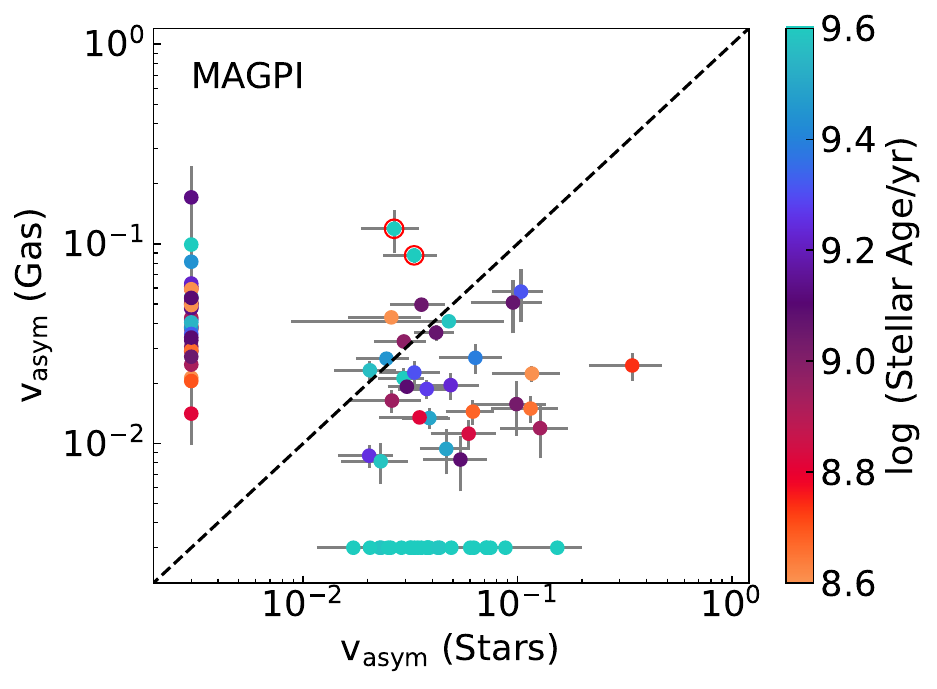}
    \caption{Stellar \vasym\ against ionised gas \vasym\ coloured by stellar age for MAGPI galaxies. Gas non-detections are shown at the bottom, stellar non-detections are shown on the left. When galaxies with old mean-stellar-ages also have gas, they tend to have larger asymmetries in their gas, rather than asymmetries in their stars. Galaxies with younger mean-stellar-ages tend to have larger stellar asymmetries, regardless if they have gas or not. Markers with red circles indicate galaxies shown in Fig. \ref{RGB}.}
    \label{asym_s_g_age_mass}
\end{figure}

\subsubsection{Gas Accretion}
\label{gas accretion}
\begin{figure}
    \centering
    \includegraphics[scale=0.4]{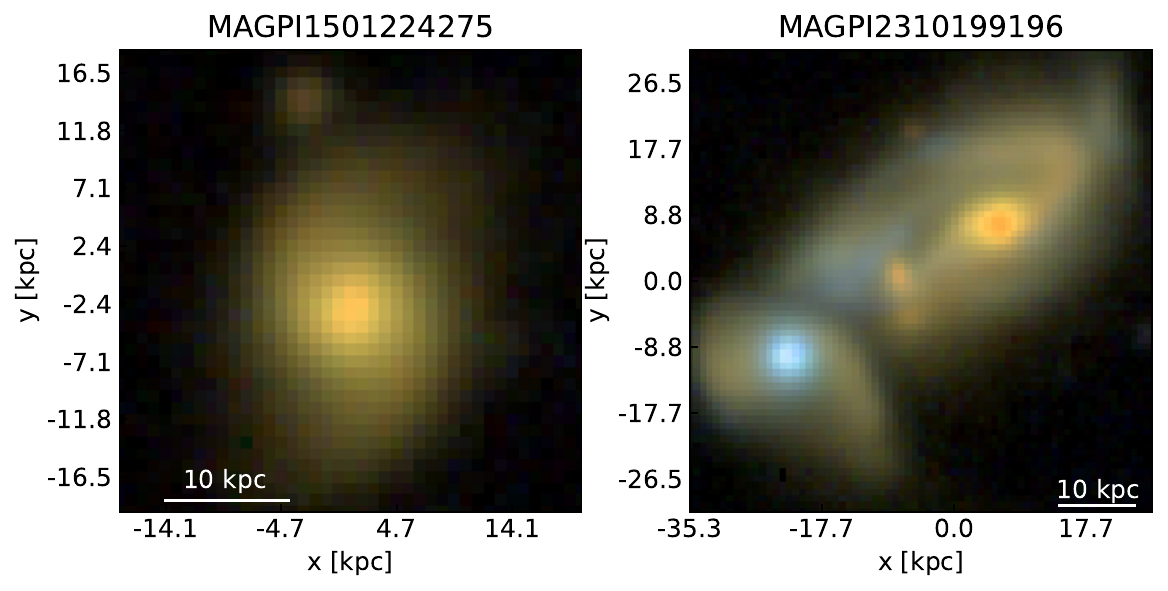}
    \caption{3 colour ($g,i,r$) images of MAGPI1501224275 (left) and MAGPI2310199196 (right). Both galaxies display larger gas asymmetries, compared to their stars, but only MAGPI2310199196 appears to be undergoing a significant merger event. We suspect MAGPI1501224275 is instead accreting gas that is independent of a merger or interaction. The faint source above MAGPI1501224275 is unconfirmed to be a nearby galaxy.}
    \label{RGB}
\end{figure}

Fig. \ref{asym_s_g_age_mass} shows the stellar asymmetries plotted against the ionised gas asymmetries for MAGPI galaxies with the points coloured by mean-stellar-age. When kinematic asymmetry is measured in galaxies with old mean-stellar-ages ($\log$ (Stellar Age/yr)>9), they typically demonstrate larger asymmetries in their gas, rather than their stars. The reverse is true for galaxies with young mean-stellar-ages, which demonstrate larger asymmetries in their stars.


Galaxies can accrete stars and gas from mergers, or they can accrete gas from the circumgalactic or intergalactic medium. Accreted gas may enter the disk with a range of angular momenta and orientation, and the dynamics of this gas will be dependent on the timescale of the accretion event. If the accretion happens over a long enough timescale where the incoming gas' angular momentum becomes aligned to the already present gas through collisions, it may not result in kinematic asymmetry. However, if the accretion event is shorter than the time it would take to realign the gas, it is likely the gas will display some kinematic asymmetry. It should also be said that this is only true for a galaxy that has an existing reservoir of gas. If a gas-poor galaxy accretes gas, the new gas would not be able to realign itself through collisions with existing gas. The accreted gas would only be able to align itself to the bulk motion of the galaxy through the torque provided by the stellar component, which would take a much longer time than if gas is present in the galaxy \citep[more than 5 dynamical times ,][]{2015MNRAS.451.3269V,2016MNRAS.457..272D}.

We considered whether the gas asymmetries in systems with old mean-stellar-ages could be due to a misaligned disk that was in the process of being rebuilt during an accretion episode \citep{2011MNRAS.417..882D,2022MNRAS.517.2677R}, however we do not find these galaxies to have large stellar-gas kinematic misalignment in their position angles ($|\Phi|<10^\circ$). Misaligned disks can be formed in galaxies that have rebuilt a disk after experiencing a major merger \citep{2006MNRAS.373..906M}. In this secnario, the incoming angular momentum of accreted gas that is misaligned is sufficiently above the torque provided by the pre-existing stellar disk, preventing the rebuilt gas disk from realigning with the stellar component \citep{2015MNRAS.451.3269V}. If accretion is responsible for the gas asymmetries, the absence of a misaligned disk would imply that the accretion rate is very low.

We postulate that the galaxies in our sample with old mean-stellar-ages, aligned gas and stellar kinematics, large gas asymmetries and low stellar asymmetries are driven by recent slow gas accretion. Accretion rather than outflows is a more likely explanation for the asymmetries in these systems since outflows occur in highly star-forming galaxies, and these old galaxies have low specific star formation rate (sSFR$<$10$^{-10}$ yr$^{-1}$). We suggest two possible scenarios for this gas accretion. Firstly, the increased gas asymmetries could arise during a gas-rich, minor merger. MAGPI2310199196 (right panel of Fig. \ref{RGB}), is one of the galaxies in our sample with old mean-stellar-ages and large gas asymmetries that is seemingly stripping the gas from the smaller, neighbouring galaxy (M$_{*,\rm neigh}$/M$_{*,\rm primary}$=0.28). Accretion of gas through minor gas-rich merger would be clumpy, explaining the elevated gas asymmetries in MAGPI2310199196. Simulations of Milky Way-like galaxies have shown that gas stripping and accretion during a minor merger can persist for up-to, and longer than 1 Gyr \citep{2018MNRAS.478.5263T,2020MNRAS.493.5636T}. This suggests that accretion of gas from a merger over an extended period could thus be clumpy enough that it causes kinematic asymmetries, but slow enough that it does not result in a kinematic misalignment.

Alternatively, the gas could been accreted externally, independent of a merger. MAGPI1501224275 (left panel of Fig. \ref{RGB}) also has large gas asymmetries compared to it stars, but it is relatively isolated and is not obviously interacting with a gas-rich galaxy. MAGPI1501224275 may instead be slowly accreting gas from the outer halo. Its old stellar population and red colour confirms that it is relatively gas-poor, and whatever little accreted gas is leading to the star formation in this system is taking longer to align with the stars due to there being no existing gas in the galaxy, resulting in larger gas kinematic asymmetries. We also considered accretion through galactic fountains; however, in this scenario, the angular momentum of the accreting gas should match that of the disk from which it was expulsed in the first place, which would not result in the asymmetries observed. Disentangling possible scenarios for the origin of high gas asymmetries in old galaxies requires careful analysis of simulations, which will be the subject of a subsequent paper.

\begin{figure}
    \centering
    \includegraphics[scale=0.5]{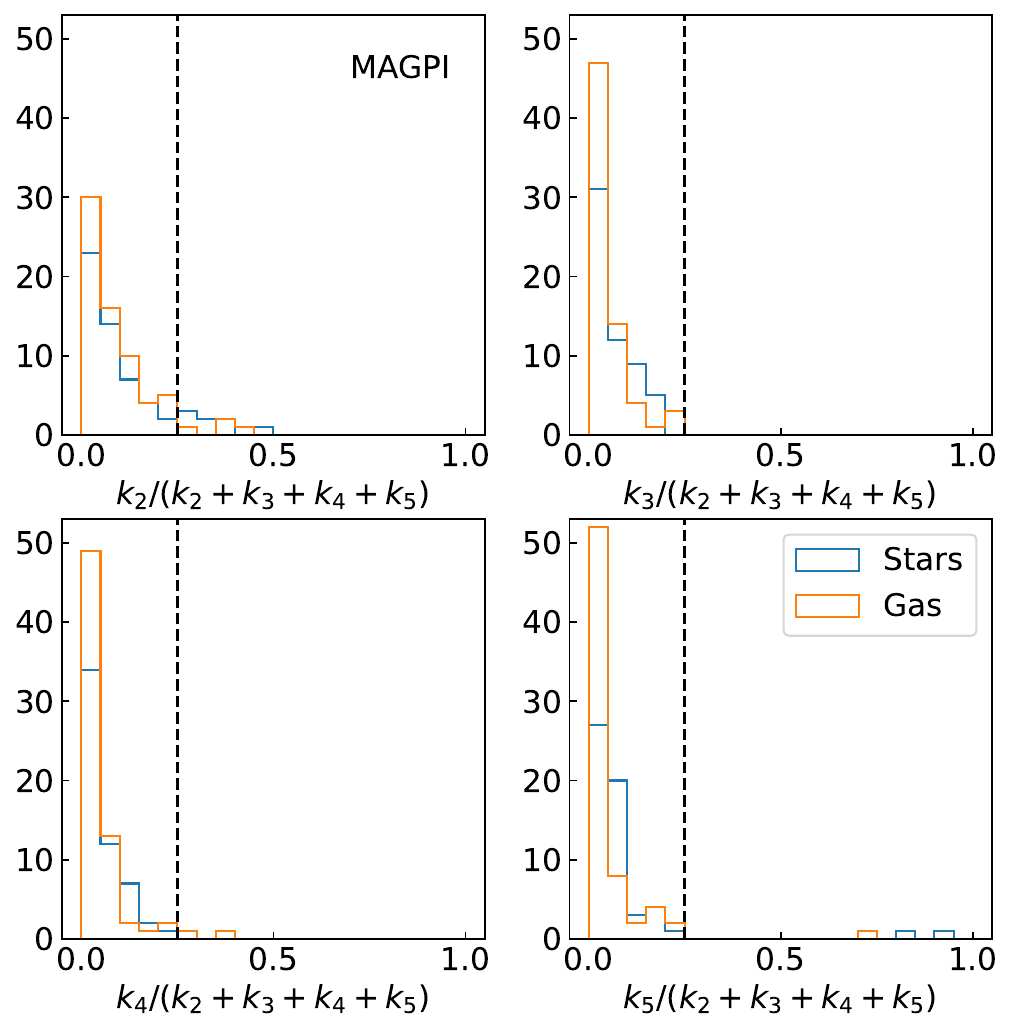}
    \caption{Histograms of $k_m$/$(k_2+k_3+k_4+k_5)$, no one mode typically ever dominates over modes in $v_{\rm asym}$. The dashed black line in each plot indicate a threshold of 0.25. No galaxies have $m=3$ as their largest mode, suggesting bar or oval-like asymmetries are not what is driving the asymmetry in our sample, given our spatial resolution.}
    \label{kn_hist}
\end{figure}

\begin{figure*}
    \centering
    \includegraphics[scale=0.7]{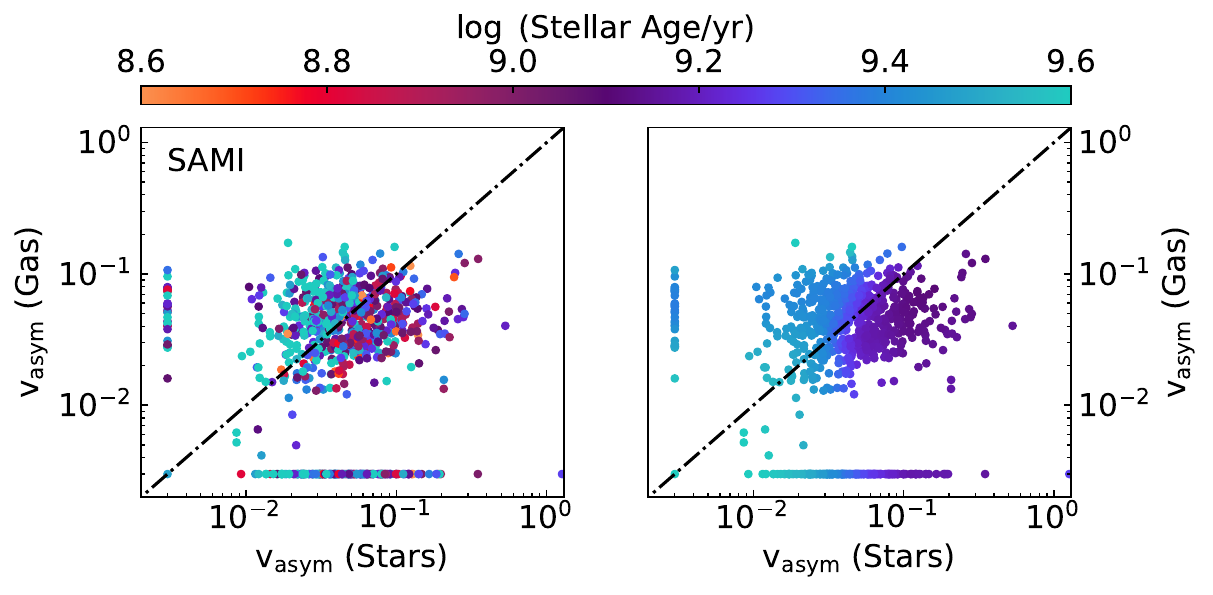}
    \caption{The same as Fig. \ref{asym_s_g_age_mass}, but for SAMI galaxies. The left panel is the raw data, the stellar ages in the right panel have been LOESS smoothed using the algorithim in \citep{2013MNRAS.432.1862C}. The trend of galaxies with old mean-stellar-ages having larger ionised gas asymmetries, and galaxies with young mean-stellar-ages having larger stellar asymmetries we found in MAGPI galaxies persists in SAMI galaxies. If accretion is responsible, this would suggest that old galaxies will continue to accrete gas over 4 Gyrs.}
    \label{SAMI_vasy_sg_age}
\end{figure*}

\begin{figure*}
    \centering
    \includegraphics[scale=0.65]{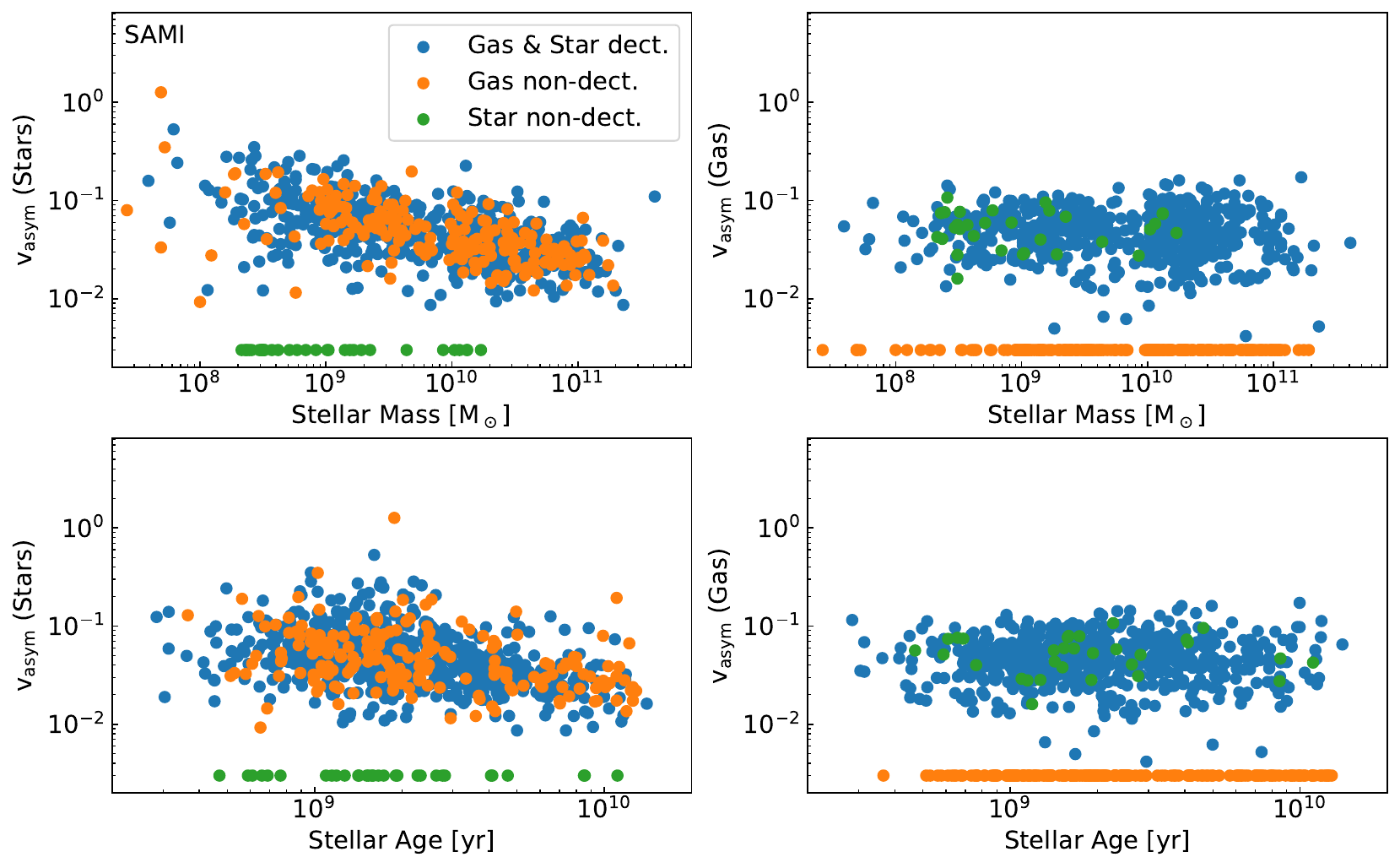}
    \caption{The same as \ref{asym_stellar_mass}, but for SAMI galaxies. We find similar trends between MAGPI and SAMI galaxies. Namely, that there is a strong correlation between stellar asymmetries and stellar mass, that is absent for the gas asymmetries. We also find that star-forming, galaxies with old mean-stellar-ages will display larger asymmetries in their gas compared to the stars.}
    \label{SAMI_vasym_age}
\end{figure*}

\subsubsection{Non-axisymmetric disk features}
Bars or `oval' distortions perturb the axisymmetric gravitational potential galaxies, which can contribute to the kinematic asymmetry \citep{1992MNRAS.259..345A, 1997MNRAS.292..349S}. Typically, this manifests as increased power to the $m=2$ and $m=3$ modes in Eqn. \ref{compactFS} \citep{2004ApJ...605..183W, 2005ApJ...634L.145G,2007ApJ...664..204S,2018MNRAS.480.2217S,2022ApJ...939...40L,2024MNRAS.527.8941T}. Fig. \ref{kn_hist} shows distributions of the individual contributions of each mode to the overall \vasym. To ascertain whether bars or oval distortions are driving kinematic assymetries in our sample, we identify galaxies where the $m=3$ mode is the dominant contributor to \vasym\ (i.e., $k_3/(k_2+k_3+k_4+k_5)>0.25$). We find no galaxy in our sample where $k_3$ is the largest contributor to the asymmetry. We repeat this process for the $k_2$ mode and find 7 galaxies where $k_2$ is the largest contributor to \vasym in either stars or gas. There are only 4 galaxies for which $k_2$ is the largest contributor for the gas $v_{\rm asym}$. Visually inspecting the images of these galaxies, we find no indication of a bar, with most having a spheroidal-like morphology. However, we note that the spatial resolution of MAGPI images is limited, which makes it difficult to identify structures such as bars (Foster et al., in prep.). Identification of bars requires sufficently resolved (2$\times$ FWHM; $\sim$6 kpc given our spatial resolution) imaging \citep{2018MNRAS.474.5372E}. While there may be hidden bars, the above exercise suggests that they are not the main driver of the kinematic asymmetries in our sample. We have confirmed that our results and conclusions are robust against removing galaxies where $k_2$ is the main contributor to the asymmetry.


\subsection{Dynamical Evolution from MAGPI to SAMI}
\label{dynamical_evolution}
\begin{figure*}
    \centering
    \includegraphics[scale=0.5]{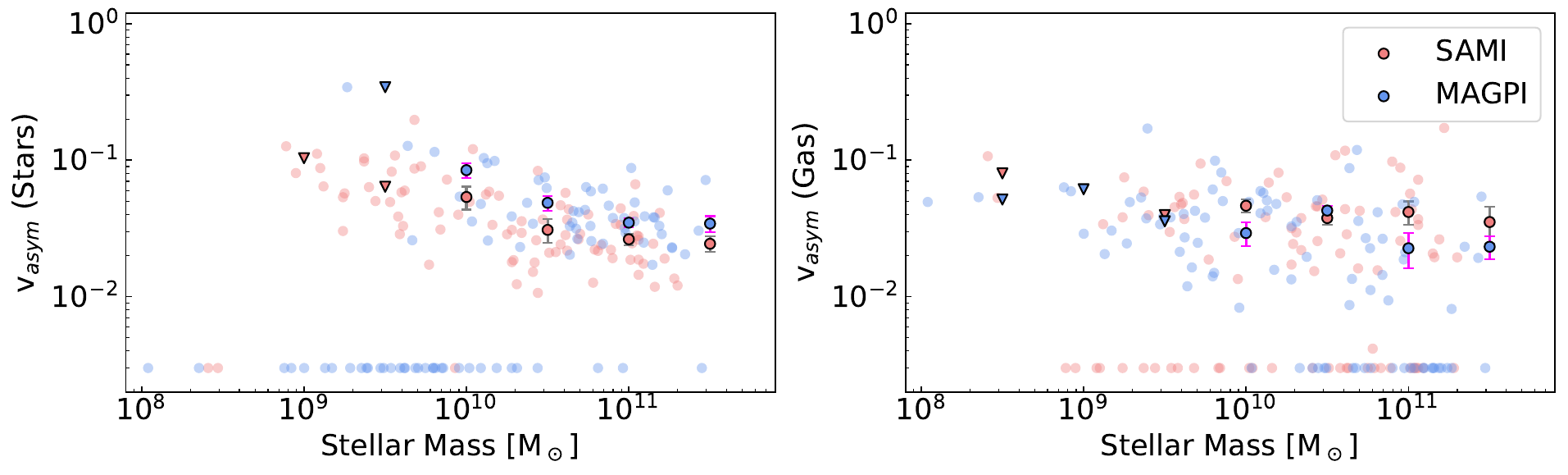}
    \caption{\textit{Left}: Stellar \vasym\ vs. stellar mass for MAGPI (blue) and SAMI (red) after stellar mass and mean-stellar-age matching the distributions. \textit{Right}: The same as \textit{left}, but for the gas asymmetries. Solid points indicate medians measured from 0.5 dex bins of stellar mass, where the errorbars represent the standard error (SE=$\sigma/\sqrt{\rm N}$). Downward facing triangles are bins with less than 10 galaxies. Non-detections are ignored when binning the data. At all stellar masses, the median \vasym\ is 1$\sigma$ higher for MAGPI galaxies compared to SAMI galaxies.}
    \label{vasym_medians}
\end{figure*}
The MAGPI survey was designed to observe galaxies at $z\sim0.3$ with similar physical resolution to local IFS surveys (MAGPI: $\sim$ 3 kpc, SAMI: $\sim$ 1.5 kpc). We can leverage the higher lookback time of MAGPI galaxies ($\sim$4 Gyrs) against SAMI galaxies ($\sim$1 Gyr) to see if the trends we are finding in MAGPI galaxies are still present in SAMI. We investigate evolution in these trends by matching the distributions in stellar mass and mean-stellar-ages between SAMI and MAGPI galaxies. While this removes the effect of stellar mass selection between surveys, it does not eliminate progenitor bias \citep{2001astro.ph..1468V,2021MNRAS.505.2247C}.

Looking at Fig. \ref{SAMI_vasy_sg_age}, we find that the trend of gas-rich galaxies with old mean-stellar-ages having larger ionised gas asymmetries compared to their stellar asymmetries is still present in SAMI galaxies. We also compare the relationship between stellar and gas asymmetries against stellar mass found in MAGPI galaxies with SAMI. Fig. \ref{SAMI_vasym_age} shows the same parameter spaces as Fig. \ref{asym_stellar_mass} but for SAMI galaxies. We find similar results in SAMI galaxies; chiefly that there is a strong anti-correlation between stellar asymmetries and stellar mass, while the correlation between gas asymmetries and stellar mass is absent. The strength of the correlation between stellar asymmetry and stellar mass is similar to what we found in MAGPI (SAMI; $\rho$=-0.58,p<10$^{-9}$, MAGPI; $\rho$=-0.59,p<10$^{-5}$). In Fig. \ref{asym_stellar_mass}, when excluding gas non-detections, we find a strong anti-correlation between mean-stellar-age and stellar \vasym\ ($\rho$=-0.49,p<0.01), though this could be driven by the larger scatter at younger mean-stellar-ages. Including the gas non-detections reduces the significance of the correlation to $\rho$=-0.28, p<0.05. Taking advantage of the larger number statistics available in SAMI, we find an anti-correlation of $\rho$=-0.36,p<10$^{-9}$. These results are consistent with low stellar mass galaxies with young mean-stellar-ages being less dynamically relaxed. The top-right panel Fig. \ref{asym_stellar_mass} also shows that, at fixed stellar mass, gas asymmetries tend to be larger in galaxies without stellar detections, however this is not apparent for SAMI galaxies. We suspect this is most likely due to low stellar mass, small, and thus low stellar continuum flux MAGPI galaxies not surviving our quality cuts/selection criteria.


From Fig. \ref{vasym_medians}, after mass and mean-stellar-age matching MAGPI and SAMI, we find that for all stellar masses, the median stellar \vasym\ is $1\sigma$ higher in MAGPI than in SAMI galaxies. This result is not mirrored in the gas \vasym. Our results are unchanged if we also match distributions of R$_{\rm e}$/FWHM prior to the mass and mean-stellar-age matching. The cause of stellar kinematic asymmetries is nuanced, while effectively driven by interactions \citep{2022MNRAS.515.3406M}, stellar asymmetries have also been found to have a stronger correlation with stellar surface density rather than environment in SAMI galaxies, when interacting/merging galaxies are excluded. Suggesting the internal mass distribution is also a significant driver \citep{2023ApJ...957L..12Z}. The elevated kinematic asymmetries in MAGPI galaxies could be due to a lower stellar surface density, however, \cite{2023MNRAS.522.3602D} found little evolution in the internal mass distribution between MAGPI galaxies and ATLAS$^{\rm 3D}$. Our results, however, are consistent with findings in \cite{2022MNRAS.515.3406M} and Pal et al., (in prep.) who studied the kinematic and photometric asymmetries in simulated merging galaxies, and MAGPI galaxies, respectively. Both report that kinematic asymmetries persist in post-merger remnants even after photometric asymmetries have dissipated.

\section{Conclusions}
\label{Conc}
We have conducted a study of the kinematic asymmetries in the stellar and ionised gas velocity maps for a subsample of MAGPI and SAMI galaxies to investigate the physical drivers behind kinematic asymmetries. Using \textsc{kinemetry}, we have quantified the asymmetry by normalising the higher order Fourier Coefficients in Eqn. \ref{compactFS} to $S_{05}$. Our findings are as follows:
\begin{itemize}
    \item Normalising the higher order terms using only $k_1$ (i.e., $V_{\rm rot}$) causes the \vasym\ to diverge in low mass, slow-rotating and more dispersion dominated galaxies. This effect is particularly prominent in the stellar asymmetries for low $\lambda_R$ galaxies (Fig. \ref{k1_vs_s05}). By normalising by $S_{05}=\sqrt{0.5 V_{\rm rot}^2 + \sigma^2}$ instead, we can measure kinematic asymmetries between the stars and ionised gas, in all galaxy types, regardless if it is a slow or fast rotator.
    \item 61\% of star-forming galaxies (those that are detected in ionised gas) in MAGPI galaxies have larger stellar asymmetries compared to gas asymmetries. The distributions of stellar asymmetries in galaxies with gas detections and with non-detections are similar (Fig. \ref{v_asym_sg_marginal}), suggesting that what is driving the stellar asymmetries does not remove the gas within galaxies in our sample, or that whatever drove the stellar asymmetries happened after the galaxy was quenched. When investigating individual modes ($k_2$ and $k_3$), we find no evidence that bars or non-axisymmetric disk features are driving the asymmetry in our sample (Fig. \ref{kn_hist}).
    \item We find an anti-correlation ($\rho=-0.59$, p$<$10$^{-5}$) between stellar asymmetry and stellar mass (top left panel of Fig. \ref{asym_stellar_mass}). We do not find an equivalent anti-correlation between ionised gas asymmetries and stellar mass (top right panel of Fig. \ref{asym_stellar_mass}). This is likely due to the gas asymmetries being removed independent of the restoring force provided from the gravitational potential (i.e., through collisional processes). The gas has more available paths to return to a state of low asymmetry than the stars do. Stars can only settle through gravitational processes (traced by their anti-correlation with $S_{05}$), resulting in the asymmetries lasting in the stars for longer.
    \item We find that star-forming galaxies with old stellar populations ($\log$ (Stellar Age/yr) $>$ 9) have larger gas asymmetries, compared to their stars (Fig. \ref{asym_s_g_age_mass}). We also find that these galaxies have little to no kinematic misalignment between the stellar and ionised gas, we suggest that these galaxies are experiencing slow, but sustained gas accretion. Employing simulations will help disentangle the source of the accretion.
    \item Investigating kinematic asymmetries in SAMI and MAGPI, we find consistent trends in both surveys. Firstly, that galaxies with old stellar populations have larger gas asymmetries, compared to their stars. Secondly, a similar anti-correlation between stellar asymmetry and stellar mass that is absent in gas asymmetries. After stellar mass and mean-stellar-age matching the distributions between SAMI and MAGPI, we find that at all stellar masses, MAGPI galaxies display larger stellar asymmetries compared to SAMI galaxies.
    \end{itemize}

We have shown that comparing kinematic asymmetries across baryonic phases, in this case, stars and ionised gas, has helped delineate certain physical drivers behind the disturbances. This study only considered galaxies at $z\sim0.01$ and $z\sim 0.3$ redshifts, but significant evolution is inferred to occur at higher redshifts from both simulations and observations \citep{2022MNRAS.509.4372L,2023arXiv230304157D}. Applying these methods to spectroscopic surveys at higher redshifts as well as mock observations of simulated galaxies will provide further understanding of the dynamical evolution in galaxies.

\section*{Acknowledgements}
We thank the referee for their helpful comments that prompted further exploration of our findings.
Based on observations collected at the European Organisation for Astronomical Research in the Southern Hemisphere under ESO program 1104.B-0536. We wish to thank the ESO staff, and in particular the staff at Paranal Observatory, for carrying out the MAGPI observations. 
MAGPI targets were selected from GAMA. GAMA is a joint European-Australasian project based around a spectroscopic campaign using the Anglo-Australian Telescope. GAMA is funded by the STFC (UK), the ARC (Australia), the AAO, and the participating institutions. GAMA photometry is based on observations made with ESO Telescopes at the La Silla Paranal Observatory under programme ID 179.A-2004, ID 177.A-3016. 
Part of this research was conducted by the Australian Research Council Centre of Excellence for All Sky Astrophysics in 3 Dimensions (ASTRO 3D), through project number CE170100013. 
CF is the recipient of an Australian Research Council Future Fellowship (project number FT210100168) funded by the Australian Government. CL, JTM and CF are the recipients of ARC Discovery Project DP210101945. 
LMV acknowledges support by the German Academic  
Scholarship Foundation (Studienstiftung des deutschen Volkes) and the  
Marianne-Plehn-Program of the Elite Network of Bavaria.
KG is supported by the Australian Research Council through the Discovery Early Career Researcher Award (DECRA) Fellowship (project number DE220100766) funded by the Australian Government. 
SMS acknowledges funding from the Australian Research Council (DE220100003).
Parts of this research were conducted by the Australian Research Council Centre of Excellence for All Sky Astrophysics in 3 Dimensions (ASTRO 3D), through project number CE170100013.
The SAMI Galaxy Survey is based on observations made at the Anglo-Australian Telescope. The Sydney-AAO Multi-object Integral field spectrograph (SAMI) was developed jointly by the University of Sydney and the Australian Astronomical Observatory. The SAMI input catalogue is based on data taken from the Sloan Digital Sky Survey, the GAMA Survey and the VST ATLAS Survey. The SAMI Galaxy Survey is supported by the Australian Research Council Centre of Excellence for All Sky Astrophysics in 3 Dimensions (ASTRO 3D), through project number CE170100013, the Australian Research Council Centre of Excellence for All-sky Astrophysics (CAASTRO), through project number CE110001020, and other participating institutions.
\section*{Data Availability}
The SAMI Galaxy Survey website is \url{http://sami-survey.org/}, and all data used in this work is publically available through Data Central \url{https://datacentral.org.au}. The MAGPI spectral data underlying this work are available from the ESO Science Archive Facility \url{http://archive.eso.org/cms.html}.



\bibliographystyle{mnras}
\bibliography{mnras_template} 




\appendix

\section{Spatial resolution across surveys}
\begin{figure*}
    \centering
    \includegraphics[scale=0.60]{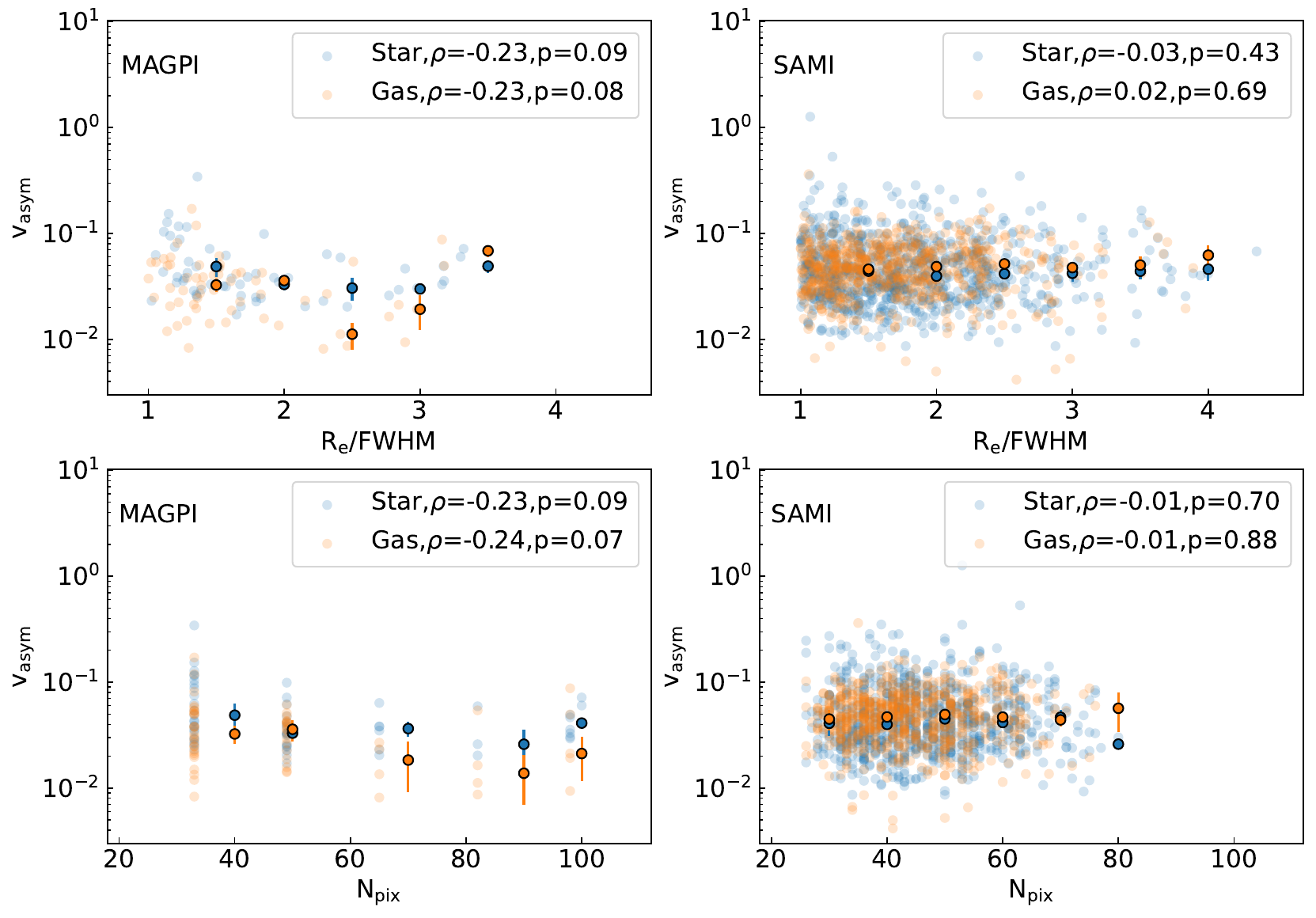}
    \caption{\textit{Top Row:} $v_{\rm asym}$ vs. $R_{\rm e}/$FWHM for MAGPI, and SAMI galaxies. \textit{Bottom Row}: $v_{\rm asym}$ vs. N$_{rm pix}$ for MAGPI, and SAMI galaxies. We find there is little correlation between how resolved a galaxy is and the asymmetry measured, for both MAGPI and SAMI samples, and both stellar and gas kinematics. Medians of binned data are shown where the error bars are the standard error from each bin. The correlation coefficents and p-values are shown in the legends of the figures. The correlation is also absent when considering number of pixels \textsc{kinemetry} uses when computing asymmetries. SAMI, with its larger number statistics would also confirms this. Given that we find little correlation between how resolved a galaxy is and the measured asymmetry, we cannot see how one would correct for seeing, or if there is even an impact.}
    \label{vasym_re_over_psf}
\end{figure*}

The median physical spatial resolution for MAGPI is roughly twice as coarse as the median physical spatial resolution for SAMI. We investigate the impact of physical spatial resoltion differences on the robustness of our results. Fig. \ref{vasym_re_over_psf} shows the asymmetry in MAGPI galaxies as a function of $R_{\rm e}/$FWHM, for MAGPI and SAMI galaxies. We find no significant trend between \vasym\ and $R_{\rm e}/$FWHM in both the gas and stars. Since \textsc{kinemetry} relies of fitting Eqn. \ref{compactFS} to pixels along ellipses, we also investigate if there is a dependency on the number of pixels sampled N$_{\rm pix}$. Reassuringly, we come to the same conclusion as with $R_{\rm e}/$FWHM. Taking advantage of the larger number statistics in SAMI, we again do not find a significant trend between R$_{\rm e}$/FWHM or N$_{\rm pix}$.

As discussed in Sect. \ref{seeing}, a correction to \vasym\ similar to that for $\lambda_{R \rm e}$ cannot be applied since we do not know the intrinsic asymmetry in the galaxy, prior to the seeing removing some of the irregularity in the velocity map. When applying a correction to $\lambda_{R\rm e}$, it can only be applied to fast-rotating galaxies, since we know \textit{a-priori} that it is rotating, and we know how seeing affects rotating systems. Since we do not have this prior assumption for kinematic asymmetry in galaxies with $R_{\rm e}$/FWHM, we cannot apply a correction. It is similar to how we can not correct slow-rotating/non-rotating systems for seeing since we do not know how beam smearing affects these systems.

\bsp	
\label{lastpage}
\end{document}